# Broadband and efficient diffraction

C. Ribot[1,2], M.S.L. Lee[2], S. Collin[3], S. Bansropun[2], P. Plouhinec[2], D. Thenot[2], S. Cassette[2], B. Loiseaux[2], P. Lalanne[1,4]

[1]Laboratoire Charles Fabry de l'Institut d'Optique, CNRS, Univ Paris-Sud, Campus Polytechnique, RD 128, 91127 Palaiseau cedex, France

[2]Thales Research & Technology, Campus Polytechnique, 1 Avenue Augustin Fresnel, 91767 Palaiseau Cedex, France

[3]Laboratoire de Photonique et de Nanostructures (LPN-CNRS), Route de Nozay, 91460 Marcoussis, France.

[4]Laboratoire Photonique, Numérique et Nanosciences, Université Bordeaux 1, CNRS, Institut d'Optique d'Aquitaine, 33405 Talence cedex, France

Contact : philippe.lalanne@institutoptique.fr

**Abstract:**
Surface topography dictates the deterministic functionality of diffraction by a surface. In order to maximize the efficiency with which a diffractive optical component, such as a grating or a diffractive lens, directs light into a chosen order of diffraction, it is necessary that it be "blazed". The efficiency of most diffractive optical components reported so far varies with the wavelength, and blazing is achieved only at a specific nominal energy, the blaze wavelength. The existence of spurious light in undesirable orders represents a severe limitation that prevents using diffractive components in broadband systems. Here we experimentally demonstrate that broadband blazing over almost one octave can be achieved by combining advanced optical design strategies and artificial dielectric materials that offer dispersion chromatism much stronger than those of conventional bulk materials. The possibility of maintaining an efficient funneling of the energy into a specific order over a broad spectral range may empower advanced research to achieve greater control over the propagation of light, leading to more compact, efficient and versatile optical components.

The basic principles for tailoring the electromagnetic field of free-space optical waves and the physics of reflection, refraction and diffraction on which it is based, have been well understood for a very long time. However, until relatively recently the whole of optical technology has been limited by the very reasonable constraint that optical systems can only be designed to be made from materials that are actually available. This paradigm is presently challenged with the introduction of electromagnetic metamaterials [1], which are artificially engineered structures that have effective dielectric or magnetic constitutive parameters not attainable with naturally occurring materials. Here we study graded metamaterial films composed of mesoscopic holes and pillars etched in a semiconductor substrate. Under illumination by an incident light whose wavelength is slightly larger than the indentation dimensions, the film implements unusually strong chromatic dispersions of the effective refractive-indices. We suggest that the fusion of low-cost lithographic techniques with this new design strategy will play a central role in devising diffractive optical components that remain blazed over a broad range of energies. Our first generation device, fabricated with fully standard semiconductor processes in a GaAs wafer, offers an efficient funnelling of the incident energy into a single diffraction orders over almost an octave.

The common way to achieve perfect blazing into a specific diffraction order is to imprint the incident electromagnetic field with a gradual phase variation that varies by exactly $2\pi$ from one side of a diffractive zone to the other. Echelette profiles with triangular grooves are common examples which have been manufactured for the last century [2]. The $2\pi$-phase jump is in general achieved at a single energy, the blaze wavelength. Illuminated at another wavelength, the $2\pi$-phase jump is not maintained and light is no longer funneled into a single order; the deviation results in the apparition of deterministic scattering into spurious diffraction orders, which represents a severe limitation for optical systems designed to operate over finite spectral bands [3]. This drawback is observed for most diffractive optical components reported so far, including classical optical diffractive components such as échelette components, spatial light modulators or volume holograms [2]. The limitation was not removed by the series of works initiated in the mid 90's, which used effective medium theory to achieve high performance blazing at visible and near-infrared wavelengths with graded-index artificial dielectric materials [4-9]. Even in recent works on metamaterial phase holograms [10] or ultra-thin metasurfaces [11-18], a broadband and efficient blazing is difficult to reach, despite the broad response of metallic nanoantennas used in these new approaches. An example of this is the light bending reported in [12], which is nicely observed over a pretty broad spectral interval from 1.1 µm to 1.4 µm, but with a low efficiency of a few percents only. In this work, a careful design that exploits the highly chromatic behavior of semiconductor structures with mesoscopic dimensions only slightly smaller than the wavelength, allows us to remove this important limitation and to achieve blazing over almost one octave in the thermal infrared. In addition since only dielectric lossless materials are involved in the structures, the blazing behavior can be made very efficient.

Figure 1 shows the grating structure. The latter is fabricated on 2-inch 300-µm-thick GaAs wafer. The wafer is first coated by an intermediate 200-nm thick silica layer deposited with a plasma enhanced chemical vapour deposition to be used as a hard mask for inductively coupled plasma etching of GaAs. To demonstrate the possibility to manufacture our diffractive optical elements at low prices and over large areas, the grating pattern is intentionally defined by classic photolithography. The grating periodicity is $a$ = 140 µm. Each period is composed of 50 different rows of subwavelength indentations, including 14 pillars and 36 holes. Fluorine-based reactive ion etching is used to transfer the resist mask into a silica hard mask. The semiconductor etching is then achieved by high-density inductively coupled plasma etching. The sacrificial $SiO_2$ layer is removed by a HF-based selective wet etching. The area of the fabricated grating is 1 cm² but all the fabrication steps are compatible with much larger processing areas.

The grating was designed for operation at a nominal wavelength $\lambda_0$ = 9.5 µm, right in the centre of the second infrared band. For this wavelength, the pillars and holes are subwavelength in size, and as the wave propagates through the diffractive elements, it experiences an artificial material with an effective index $n_{eff}$ shown in Fig. 2a. The effective indices have been calculated by assuming that the indentations are regularly placed on a square grid with a 2.8µm-periodicity [19]. The tiny air holes correspond to a dense artificial dielectric with a large effective index close to the refractive index of the bulk material ($n_M$ = 3.30) and the smaller GaAs pillars with a 0.64 fillfactor correspond to a material with an effective index of $n_m$ = 1.90. For the design, the 50 indentations of the period are chosen to achieve a linear gradient of the effective index. Thus for an etching depth $d = \lambda_0/(n_M − n_m)$, a gradual phase variation from 0 to $2\pi$ from one side of the period to the other is introduced and the grating is blazed in transmission at $\lambda_0$, implying that all the transmitted energy is scattered into a single diffraction order. Actually the present device is nearly perfectly blazed, since ∼90% of the transmitted energy is funneled into a dominant diffraction order. Nearly perfect blazing has already been achieved with graded-index artificial dielectric materials [7] at visible and infrared energies, and strictly speaking, if we restrict ourselves to single frequency operation, nothing is really novel at this stage. Indeed the key point is the spectral properties. Actually blazing persistently remains as the energy of the incident beam is tuned from the nominal energy.

Broadband blazing is qualitatively evidenced by the grey-scale transmittance diagram $T(\lambda,\theta)$ shown in Fig. 3. The data are recorded for a TE-polarized plane-wave illumination at normal incidence from the flat-side of the wafer (see the bottom-right inset) in the spectral range 3−17 µm and for various detection angles $\theta$ ranging from −15° to 11° with a step of 0.1°. Measurements are performed with a set-up based on a commercial Fourier transform spectrometer (FTIR) and on a home-made achromatic imaging system enabling well-collimated beams (divergence angle of 0.5°) with a spot diameter of 1.7 mm. The spectral resolution is set to 20 cm$^{-1}$. Details of the set-up can be found in [20]. The dominant branch in the diagram corresponds to the negative-first order, $\sin(\theta) = m\lambda/a$ with $m = −1$. It captures most of the transmitted energy, a small fraction of the transmitted light being also carried out into other diffraction orders, especially the zeroth order, as evidenced by the saturated grey-scale miniature diagram shown on the top-right side. Remarkably, this is achieved over a broad spectral interval, qualitatively from 5 µm to 15 µm.

The peculiarity of the present device is the use of two different subwavelength patterns for achieving the $2\pi$-phase variation across the grating periods. The combination relaxes the fabrication constraints on the etch depth and allows us to fully exploit the highly dispersive nature of artificial dielectrics with mesoscopic dimensions to render the $2\pi$-phase variation almost achromatic. To realize that in an intuitive description, let us first consider what happen when the phase variation is implemented with achromatic materials. This occurs in practice for indentations with deep-subwavelength dimensions operating in the homogeneization regime [4] or for échelette elements for which the bulk material dispersion is negligible. For such components, the phase difference $\Delta\Phi$ seen by two waves sensing the two sides of any diffractive zone is $2\pi$ but only at the nominal wavelength $\lambda_0$, as $\Delta\Phi$ varies linearly with the frequency, $\Delta\Phi = 2\pi\lambda_0/\lambda$. The wavelength-dependence of $\Delta\Phi$ for such achromatic-material diffractive elements is shown with the solid blue curve in Fig. 2c. The net consequence is that such devices are blazed only at the nominal wavelength $\lambda_0$, and as one departs from $\lambda_0$, the efficiency drops: light is scattered into spurious orders.

The present design seriously challenges this classical limitation by carefully exploiting the highly dispersive nature of mesoscopic metamaterials [21-24]. Figure 2b shows the frequency dependence of the maximum and minimum effective indices of the metamaterials used at the extremities of the grating periods. The maximum value $n_M$ (solid curve) is achieved for "null-diameter holes" (bulk GaAs) and is nearly achromatic. In contrast the minimum value $n_m$ (dashed curve) achieved for narrow pillars ($f = 0.64$) varies with the wavelength, except for wavelengths that are substantially larger than the pillar dimensions where the homogeneization regime is reached. On overall, the effective-index difference $n_M-n_m$ is highly chromatic, and as it monotonously increases with the wavelength, the phase difference $\Delta\Phi = (2\pi/\lambda)(n_M-n_m)d$, albeit not fully achromatic, is weakly dependent of the wavelength. $\Delta\Phi$ is shown with the dashed curve in Fig. 2c. Noticeably, it presents an extremum at the nominal wavelength and a spectral behavior that is much flatter than that achieved with non-dispersive materials (solid blue curve). It is perhaps worth emphasizing that it is the low effective index value $n_m$ that disperses in the present design. $n_m$ decreases with the wavelength (like for most common natural bulk material), and therefore $(n_M-n_m)$ increases, letting $(2\pi/\lambda)(n_M-n_m)$ weakly dependent on the wavelength. In earlier designs of graded-index artificial dielectric materials based on pillars only, see Refs. [6-9] for instance, it is actually the large effective index value $n_M$ that disperses, and therefore $(n_M-n_m)$ decreases with the wavelength. The wavelength-dependence of $\Delta\Phi$ is accentuated and the spectral behavior is even worse than that of classical diffractive elements made of achromatic materials.

Figure 4 shows the diffraction efficiencies of the 0th, $-1$st and $-2^{nd}$ orders for illumination at normal incidence. The left scale represents absolute efficiencies, $E_a$, defined as the ratio between the energy funneled into a specific order and the incident energy impinging on the diffractive element (measured by simply removing the sample). Absolute efficiencies include reflection losses at the rear and front surfaces of the wafer. Because of the large refractive index of GaAs ($n_{GaAs} \sim 3.3$), back-reflection is substantial. The right scale is corrected for the reflection loss ($\sim 49\%$ over the spectral range of interest) incurred by two flat air-GaAs interfaces. It represents the relative efficiency $E_r = E_a/0.51$, defined as the ratio between the energy funneled into a specific order normalized by the specular transmission of the GaAs wafer illuminated outside the patterned region. Note that $E_r$ is not bounded to 100% since the transmittance of the patterned rear-interface may be larger than that of the flat interface, depending on the wavelength. The salient result of Fig. 4 is evidenced by the blue rectangular window. The later shows that the incident energy is dominantly funneled into the $-1^{st}$ order over almost one octave, from 8 µm to 15 µm. On spectral average, the relative efficiency of the $-1^{st}$ order is 90%, whereas the relative efficiencies of the 0th and $-2^{nd}$ orders are smaller than 6% and 3%, respectively. About 85% of the total transmitted light is funneled into the desired order. We have additionally checked that the efficiencies weakly depends on the polarization. The dotted curves represent theoretical predictions of the efficiencies. They are obtained with a semi-analytical method that combines scalar and vector-theory [25]. The method takes into consideration many fabrication imperfections, such as the actual measured size of the cylinders and pillars, which differs from the design values and the fact that the etching depth varies from 4.3 µm for tiny holes to 7 µm for large holes and pillars. The theoretical predictions confirm the realization of broadband blazing. We note however that the $-1^{st}$ order experimental values are lower than the theoretical ones. The systematic offset may be explained by the rounded (not square) shape of the fabricated indentations that is not considered in the theoretical model and to an inevitable 2% scattering loss induced by the geometry rupture at the pillar-hole transition (see Fig. 1).

In conclusion, the results reported herein demonstrate the possibility of achieving broadband and efficient diffraction, a fundamental limitation common to all diffractive elements fabricated so far. This was achieved with carefully designed artificial dielectric surfaces for the textbook case of a first-order blazed grating, but we emphasize that any arbitrary phase function can be designed with the same design rules and manufactured with the same technological process.

Presently the device efficiency (absolute efficiency of 50%) is limited by impedance mismatch. The reflection at the rear flat interface ($\approx$ 30%) is not the limiting point. More problematical is the front etched surface. Graded-index gratings manufactured in bi-layered artificial dielectrics have already shown good antireflection capabilities at the nominal blazed wavelength [26], but the bi-layer approach is likely not to be broadband enough. High-performance broadband antireflection coats of semiconductors are naturally achieved with pyramidal structures [27], but it is not clear how this approach may be easily implemented in the present context without requiring very high aspect ratio that are difficult to manufacture.

The possibility of manufacturing diffractive elements with spectrally-flat efficiencies, with efficiencies as high as possible, is very likely to be a valuable property for many applications. Such a challenge should be taken up by the metamaterial community. This work is an example of a new strategy. The approach can be generalized to other regions of the electromagnetic spectrum, even in the visible or near-infrared with semiconductors that provide very large index-contrast and thus very-high artificial chromatism, provided that some residual absorption is acceptable. Absorption does not prevent blazing [9], and for many applications, loss is not the main issue, especially if the energy scattered into the spurious orders is maintained at a very low level, like in the present work. Perhaps ultra-thin metasurfaces manufactured with metal resonances [11-18] may open new avenues for low cost ultra-braodband, since the approach does not rely on high-aspect ratio subwavelength structures.

**Acknowledgements**
The authors acknowledge partial financial support form the Délégation Génération de l'Armement.  The research has also been supported in part by the ANR project MICPHIR and the EMRS-DTC programme. J.P. Huignard, Christophe Sauvan and Pierre Chavel are gratefully acknowledged for fruitful discussions.

**Figures & figure captions**

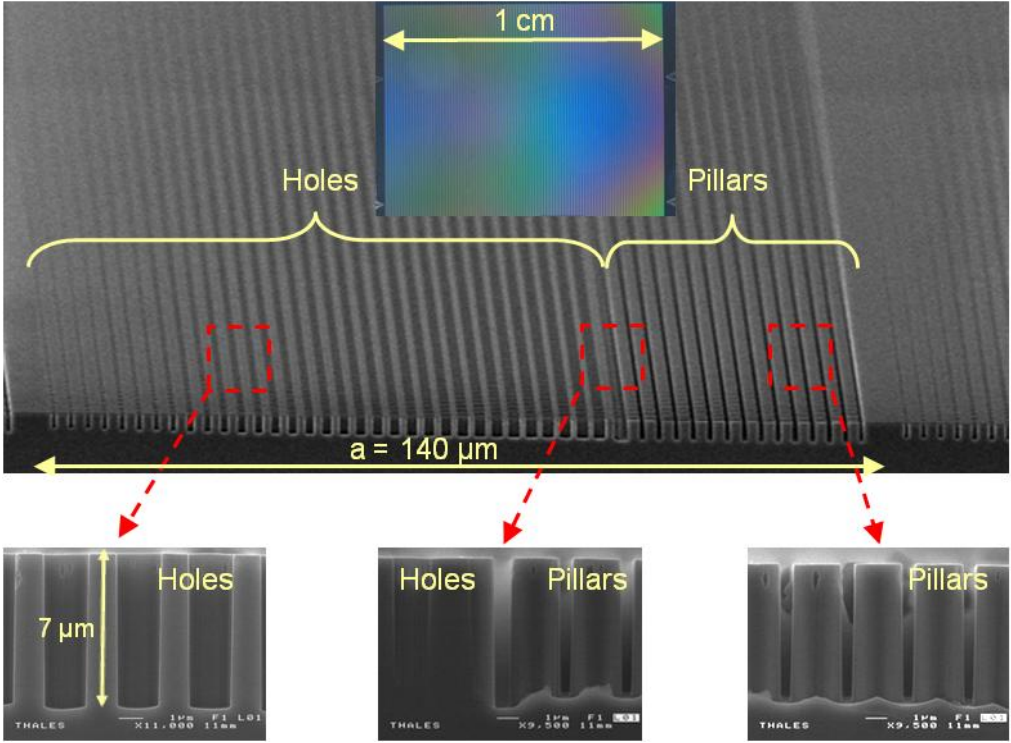

Figure 1. **Scanning electron micrograph picture showing a single period of the grating.** The period is 140-µm long and is composed of 50 subwavelength pillars and holes etched in a GaAs wafer with progressively varying dimensions. The effective refractive-index decreases linearly from left to right. The total area is 1 cm² (see the optical photograph of the top inset).

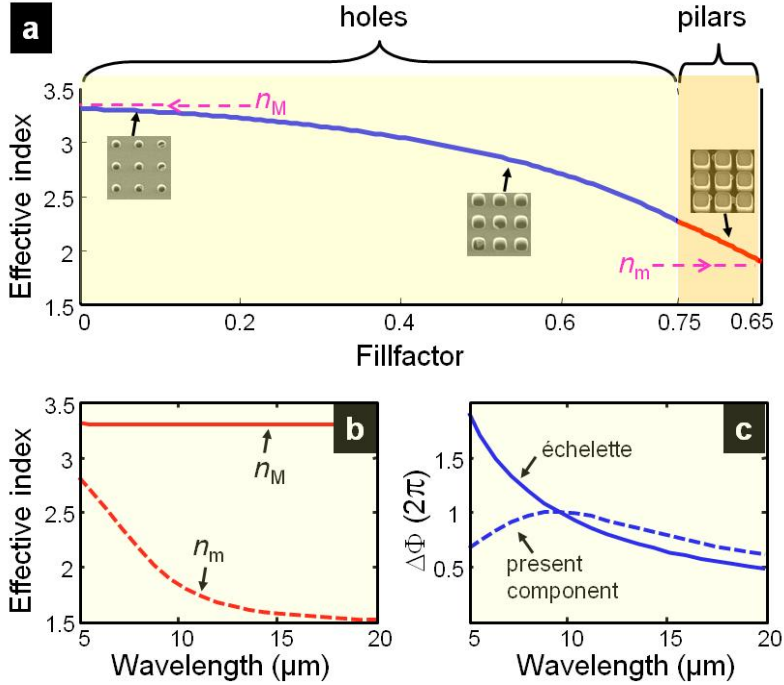

Figure 2. **Effective dielectric constitutive parameters of the pillar and hole structures.** **(a)** The effective refractive index is presented at a nominal wavelength $\lambda_0 = 9.5$ μm for beam propagating parallel to the pillar and hole axes. It is independent of the polarization. The fillfactor *f* is defined as the ratio of the square pillar (or hole) width to the indentation pitch. The maximum effective refractive index $n_M = 3.30$, encoded with holes with vanishingly-small fillfactors, corresponds to the refractive index of the bulk GaAs wafer. The smaller index $n_m = 1.90$ is achieved with an artificial dielectric composed of pillars with a small 0.64 fillfactor. **(b)** Chromatic dispersion properties of $n_m$ and $n_M$. Only the effective dielectric composed of tiny pillars is a highly dispersive medium. **(c)** Wavelength-dependence of the phase variation $\Delta\Phi = (2\pi/\lambda)\,(n_M - n_m)d$ across the grating periods. The solid curve corresponds to diffractive elements made of bulk (almost dispersion-less) materials.

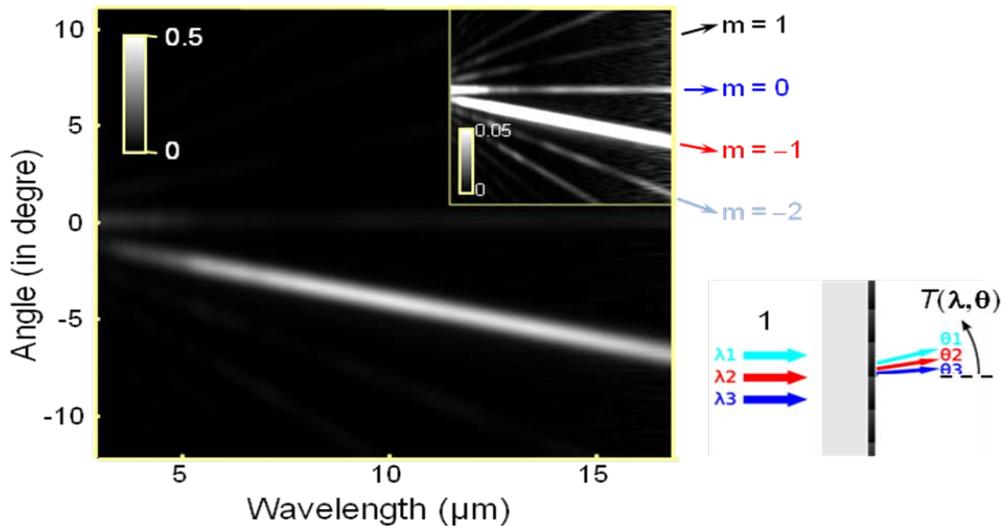

Figure 3. **Grey-scale transmittance diagram $T(\lambda,\theta)$ evidencing broadband blazing,** where $\theta$ represents the detection angle. At a given wavelength, light is diffracted into specific diffraction angles given by the grating equation: $\sin(\theta) = m\lambda/a$. The transmitted power is funneled into a dominant order, the $-1^{st}$ order, which carries most of the transmitted energy (see the miniature inset that reveals the excitation of spurious orders with a saturated grey-scale). The data are recorded for a TE-polarized plane-wave illumination incident from the flat-side of the wafer at normal incidence (see the left inset) with a home-made FTIR spectrometer.

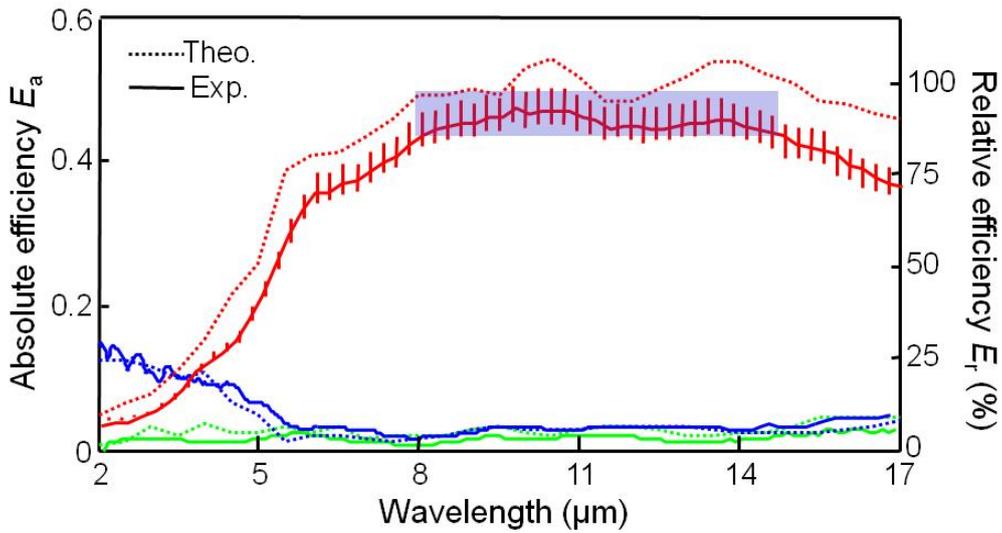

Figure 4. **Quantitative analysis of the broadband funneling.** Absolute and relative efficiencies (see text for a definition) for unpolarized light are displayed on the left and right vertical scales for the dominant −1st order (red) and for the 0th (blue) and −2nd (green) orders. Experimental and theoretical data are shown with solid and dotted lines, respectively. The vertical red bars represent uncertainties due to systematic and random errors, estimated from measurements performed for both polarizations.